\documentclass[12pt,a4paper]{article}
\usepackage{a4wide}

\usepackage{latexsym,amssymb,amsmath,epsfig,amsfonts,algorithm}
\usepackage{multirow,multicol,array}
\usepackage{amssymb}
\usepackage{nccmath, mathtools}
\usepackage{cite}
\usepackage[colorlinks=true,allcolors=blue]{hyperref}
\usepackage{float}
\usepackage[caption = false]{subfig}
\usepackage{caption}
\usepackage[utf8]{inputenc}
\usepackage[english]{babel}
\usepackage[noend]{algpseudocode}
\usepackage{tikz}
\usetikzlibrary{arrows}
\usepackage{bbm}
\usepackage[noend]{algpseudocode}
\newcommand\numberthis{\addtocounter{equation}{1}\tag{\theequation}}

\newcommand{\ds}{\displaystyle}

\DeclareMathOperator{\EX}{\mathbb{E}}

\renewcommand{\theequation}{\arabic{section}.\arabic{equation}}

\providecommand{\keywords}[1]{\textbf{\textit{Key words---}} #1}
  
\newtheorem{theorem}{Theorem}[section]

\numberwithin{figure}{section}
\numberwithin{table}{section}
\numberwithin{equation}{section}

\begin{document}
\title{The Value of Communication and Cooperation in a Two-Server Service System}

\author{Mark Fackrell, Cong Li, Peter Taylor, Jiesen Wang\\
	\normalsize{School of Mathematics and Statistics, The University of Melbourne,}\\
	\normalsize{Victoria 3010, Australia}\\
	\normalsize{contact:\{fackrell, li.c, taylorpg\}@unimelb.edu.au, jiesenw@student.unimelb.edu.au}
}
\date{}
\maketitle
\begin{abstract}
	In 2015, Guglielmi and Badia discussed optimal strategies in a particular type of service system with two strategic servers. In their setup, each server can either be active or inactive and an active server can be  requested to transmit a sequence of packets. The servers have varying probabilities of successfully transmitting when they are active, and both servers receive a unit reward if the sequence of packets is transmitted successfully. Guglielmi and Badia provided an analysis of optimal strategies in four scenarios: where each server does not know the other's successful transmission probability; one of the two servers is always inactive; each server knows the other's successful transmission probability; and they are willing to cooperate.

	Unfortunately the analysis in Guglielmi and Badia contained errors. In this paper we correct these errors. We discuss three cases where both servers (I) communicate and cooperate; (II) neither communicate nor cooperate; (III) communicate but do not cooperate. In particular, we obtain the unique Nash equilibrium strategy in Case II through a Bayesian game formulation, and demonstrate that there is a region in the parameter space where there are multiple Nash equilibria in Case III. We also quantify the value of communication or cooperation by comparing the social welfare in the three cases, and propose possible regulations to make the Nash equilibrium strategy the socially optimal strategy for both Cases II and III.
\end{abstract}


\maketitle

\section{Introduction} 
\noindent Numerous workplaces require servers to complete crucial and highly technical tasks. These tasks may be subject to service failure and incur operational fees, so a server may not be willing to complete them. For example, controlling the spread of highly infectious diseases or managing the disposal of radioactive waste always demands highly skilled staff, and there is the possibility of service failure. The system manager aims to accomplish as much work as possible while minimising the overall operating cost, in a situation where servers have a tendency to avoid a job being allocated to them and to let their co-workers undertake the task. The situation is common in 
a network layer with distributed routing selections \cite{MZRR03}, a data link layer with multiple access techniques \cite{JBGW10}, and devices powered by distributed energy sources \cite{MMZ12}.  Game theory, which has already been applied to study telecommunication problems 
\cite{HNDB12}, provides a mathematical framework with which to model and analyse this situation since it considers the behaviour of individuals given that everyone's best strategy is affected by the strategies adopted by other participants in the system. In this paper we shall take another look at the system that was studied by Guglielmi and Badia \cite{AGLB15} and derive somewhat different results.

The paper is organised as follows. In Section 2 we describe the strategic server system proposed by Guglielmi and Badia \cite{AGLB15}, and compare our analysis with theirs. In Section 3 we introduce some preliminaries on game theory and how we use them to obtain the main results. In Section 4 we derive the optimal strategy  if each server has complete knowledge of the other's probability of successful transmission $p_i$ and they are willing to cooperate. In Section 5 we obtain the unique threshold type Nash equilibrium strategy \cite[Section 1.1.6]{RHMH03} under the scenario in which each server is informed only that the other server is using a threshold strategy, and what the threshold is. In Section 6 we describe the Nash equilibrium strategy in the case when both servers know each other's probability of successful transmission $p_i$, but individuals maximise their own payoffs without considering the effect on the other server. The paper concludes with Section 7 where we summarise our results and propose some directions for future research.

\section{The strategic server system of Guglielmi and Badia}
\noindent Guglielmi and Badia \cite{AGLB15} proposed and analysed a transmission system of two servers. Each server has to make a decision whether to be {\it active}, that is available for the service, or {\it inactive}, that is not available for the service. A server who is available for the service will be chosen to undertake the service via a rule that we shall discuss below. 

For $i = 1,2,$ server $i$'s probability of successfully transmitting a packet when it is active is $p_i$. The $p_i$s are assumed to be independent and identically distributed according to a continuous uniform distribution on $[0, 1]$, and each server knows its own $p_i$. We call $(p_1, \, p_2)$ the {\it state} of the system, and so $[0, 1] \times [0, 1]$ is the {\it state space} of the system. The model assumes that a successful transmission gains a reward of 1 for each server, whether or not they were the one to undertake the service. The one-off cost of being active is $c \in [0, 1]$. If both servers are active, the service is allocated to the one with the higher probability of successful transmission. 

For $i=1,2$, we shall use $-i$ to denote server $j \neq  i$. Then the servers' expected payoffs, which depend on their decisions and ($p_1, p_2$), have four different cases. If both servers are inactive, both payoffs are 0 because the service is not completed. If server $i$ is active and server $-i$ is inactive, the expected payoff to server $i$ is $p_i - c$ and to server $-i$ is $p_i$. This is because the probability of successful transmission is $p_i$ in this case, so both servers gain a reward of 1, while server $i$ needs to pay a cost $c$ for being active; if both servers are active, both expected payoffs are $\max\{p_1,p_2\} - c$, because now the probability of successful transmission is $\max\{p_1, p_2\}$ and both servers need to pay $c$ for being active. The expected payoff matrix is shown in Table \ref{tab1}, where the expected payoffs to servers 1 and 2 are given by the first and second coordinates, respectively. 

\begin{table}
	\setlength{\extrarowheight}{2pt}
	\centering
	\begin{tabular}{cc|c|c|}
		& \multicolumn{1}{c}{} & \multicolumn{2}{c}{server $2$}\\
		& \multicolumn{1}{c}{} & \multicolumn{1}{c}{$active$}  & \multicolumn{1}{c}{$inactive$} \\\cline{3-4}
		\multirow{3}*{server $1$}  & $active$ & $(\max\{p_1,p_2\} - c, $ & $(p_1-c,p_1)$ \\
		& $ $ & $\max\{p_1,p_2\} - c)$ & $ $ \\\cline{3-4}
		& $inactive$ & $(p_2,p_2-c)$ & $(0,0)$ \\\cline{3-4}
	\end{tabular}
	\caption{Expected payoff matrix of the game} \centering\label{tab1}
\end{table}

Guglielmi and Badia \cite{AGLB15} applied game theory to analyse four scenarios. 
\vspace{1mm}
\begin{itemize}
	\item In Scenario 1, each server does not know the other's probability of successful transmission. They claimed that the Nash equilibrium strategy is of threshold type. That is, server $i$ chooses to be active when $p_i \geq p^*_i \,, i =1,2$, and the threshold values $p^*_1$ and $p^*_2$ must satisfy $p^*_1\, p^*_2 =c$.
	\\
	\item In Scenario 2, they examined two special cases of Scenario 1 where one of the two servers is always  inactive. They claimed that the two cases are the worst case allocation with respect to  social welfare. 
	\\
	\item In Scenario 3, both servers have full information about each other's $p_i$. They obtained results similar to those in Scenario 1. Each server chooses to be active only if $p_i \geq p^*_i \,, i =1,2$, but server 1 chooses to be  inactive when $p^*_1 \leq p_1 \leq \mfrac{p^*_1-1}{p^*_2-1}p_2-\mfrac{p^*_1-p^*_2}{p^*_2-1}$, and server 2 chooses to be inactive when $p^*_2 \leq p_2 \leq \mfrac{p^*_2-1}{p^*_1-1}p_1-\mfrac{p^*_2-p^*_1}{p^*_1-1}$.
	\\
	\item In Scenario 4, coordinated servers have complete knowledge about each other's $p_i$. They derived the unique Nash equilibrium strategy which is of threshold type with threshold value $c$. That is, each server chooses to be active if and only if $p_i \geq c \,, i =1,2$. 
\end{itemize}
\vspace{1mm}
Unfortunately, the analysis of \cite{AGLB15} contained errors. In this paper, we correct the errors and change the order so that we look at the most straightforward case first:
\vspace{1mm}
\begin{itemize}
	\item In Case I, which corresponds to Scenario 4, both servers have full knowledge of the other's probability of successful transmission $p_i$ and both try to optimise social welfare. The best strategy is of threshold type, but the threshold value is $\mfrac{\ds c}{\ds 2}$ instead of $c$. That is, each server chooses to be active if and only if $p_i \geq \mfrac{\ds c}{\ds 2} \,, i =1,2$.
	\\
	\item In Case II, which corresponds to Scenario 1, each server does not know the other's probability of successful transmission. The threshold strategy with threshold value $\sqrt{c}$ is a Nash equilibrium strategy. That is, if both servers choose to be active when $p_i \geq \sqrt{c} \,, i=1,2$, neither server has an incentive to deviate. In addition, we prove this is the only threshold Nash equilibrium strategy. This is different from the conclusion stated by Guglielmi and Badia \cite{AGLB15} that all pairs satisfying $p^*_1\, p^*_2 =c$ are Nash equilibria. Also, we calculate the expression for social welfare for the threshold strategies and obtain the best case allocation. It is clear from the expression that the worst case is not necessarily when one of the two servers is always inactive as stated in Scenario 2.
	\\
	\item In Case III, which corresponds to Scenario 3, each server knows the other's probability of successful transmission. We show that server $i$ chooses to be active only if $p_i \geq c \,, i =1,2$, and there is a region in the parameter space where there are multiple Nash equilibria. The regions where each server is active in our analysis is different from that in \cite{AGLB15}, and we also obtain a mixed Nash equilibrium strategy in a specific region.
	\\
	\item For both Case II and Case III,  we propose regulations and prove that by imposing these regulations, we can eliminate the social inefficiency caused by noncooperation.
\end{itemize}

\section{Preliminaries}
In game theory, an action profile $(a_1, a_2)$ represents the actions adopted by both servers, and it could be pure or mixed. If, for $i = 1,2$, we let $A_i = \{active,  inactive\}$, then for server $i$, a pure strategy is $a_i\in A_i$,  and a mixed strategy is $a_i = \sigma_i$ where $\sigma_i$ is the probability that server $i$ is active. A Nash equilibrium is an action profile where no server has an incentive to deviate unilaterally. We use $s = (s_1, s_2)$ to denote the strategy on the whole state space, and $BR_i(s_{-i})$ to denote server $i$'s best response given server $-i$ adopts strategy $s_{-i}$. In many queueing models, $s_i$ can be represented by a single numerical value \cite[Section 1.1.6]{RHMH03}. In this paper, we let $(\underline{p^*_1},\underline{p^*_2})$ denote a threshold strategy where server 1 is active if and only if $p_1 \geq p^*_1$, and server 2 is active if and only if $p_2 \geq p^*_2$, where $0 \leq p^*_1,\,p^*_2 \leq 1$. In Section 4, we use $(a_1, \underline{p^*_2})$ to represent the situation where server $1$ takes action $a_1$ and server $2$ adopts the threshold strategy with threshold value $p^*_2$. For a given strategy $s$, we define the {\it social welfare} $S_O(s)$ to be the expected value of the sum of the two servers' payoffs. We define the social optimum as the maximum social welfare under certain assumptions. 

A system that is in a Nash equilibrium is not necessarily at a social optimum, but we can sometimes regulate the game to adjust a Nash equilibrium to match the strategy of the optimal social welfare. For example, a queue could be regulated by announcing an admission fee \cite{PN69} or imposing a toll on waiting \cite[p24]{RHMH03}. Li \cite{CL17} generalised the setup in Guglielmi and Badia \cite{AGLB15} by introducing an incentive parameter $b$ to make the payment system fairer. When a service is completed,  the server who performs the service is awarded 1, and the server who does not perform the service is awarded $b \, ( \leq 1)$. So the setup of Guglielmi and Badia \cite{AGLB15} had $b=1$. Li \cite{CL17} showed that by introducing $b$, the servers are encouraged to stay in the system.

We apply game theory to analyse this model. More specifically, we employ {\it Bayesian game theory} \cite[Section 2.6]{MOAR94} to examine Case II where both servers are not informed of the other's probability of successful transmission, but each is aware that the other is adopting a threshold strategy and what the threshold value is. That is, for $i = 1,2$, server $i$ does not know $p_{-i}$, but knows server $-i$ is  active if and only if $p_{-i} > p^*_{-i}$ and what $p^*_{-i}$ is. A Bayesian game is designed to analyse the situation with imperfect information and models servers' information about the state of nature by prior belief and its type. The type is the signal that each server observes. We let $p_i$ denote server $i$'s type, and $U_i((a_i, \underline{p^{*}_{-i}}), p_i)$ denote its payoff. That is, server $i$'s payoff depends on its type $p_i$ and the strategy $(a_i, \underline{p^{*}_{-i}})$. We prove that in this situation, the Nash equilibrium strategy is unique and also of threshold type. For the case in which both servers have full information about each other but act selfishly, the signal for each server is the same, that is, each server knows its own $p_i$ and the other server's $p_{-i}$. We show that a Nash equilibrium strategy is not necessarily of threshold type in the full information case. We also propose regulations and prove that by imposing these regulations, we can eliminate the social inefficiency caused by noncooperation.


\section{Case \uppercase\expandafter{\romannumeral1}: Cooperative servers with communication}

Case \uppercase\expandafter{\romannumeral1} represents the situation in which both servers have full knowledge of the other's probability of successful transmission $p_i$ and both try to optimise social welfare. This models the perspective of the system manager, and the social welfare in this case is maximal under the environment described in Section 1.  The best strategy for both servers to maximise social welfare is described in Theorem \ref{thm1}.
\\[1pt]
\begin{figure} 
	\centering
	\includegraphics[width = .5\textwidth]{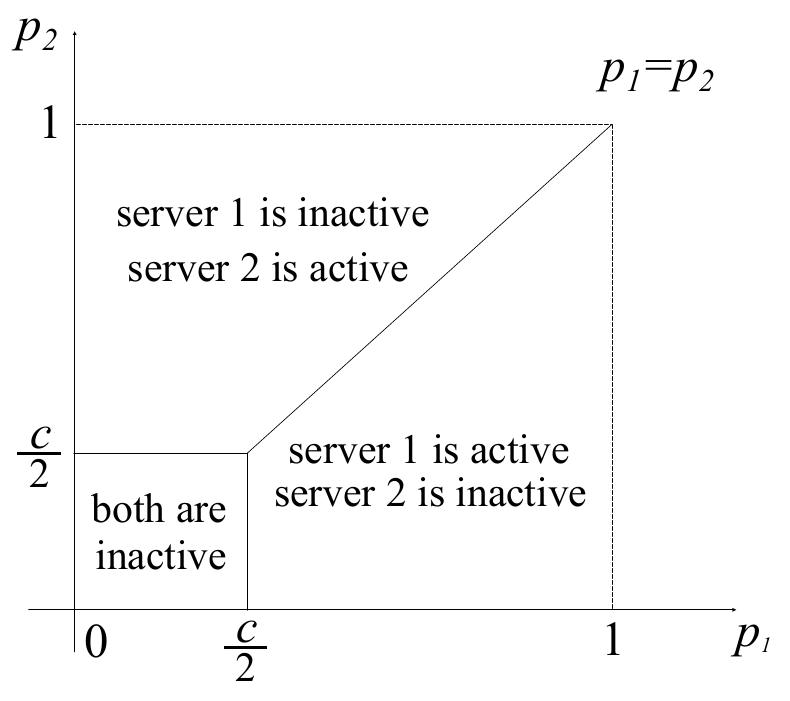}
	\caption{Optimal action profile for cooperative servers with communication} \centering\label{1}
\end{figure}

\begin{theorem} \label{thm1}
	If $p = (p_1, p_2)$ is observable to both servers when they make decisions, the best strategy to maximise overall welfare is 
	\begin{equation}\label{s*}
	s^* =
	\begin{cases}
	({\it inactive}, {\it inactive}), & \max\{p_1, p_2\} < \frac{c}{2} \\
	(\sigma_1, {\it inactive}), & \max\{p_1, p_2\} = p_1 = \frac{c}{2} \\
	({\it inactive}, \sigma_2), & \max\{p_1, p_2\} = p_2 = \frac{c}{2} \\
	({\it active}, {\it inactive}), &  \max\{p_1, p_2\} = p_1 > \frac{c}{2} \\
	({\it inactive}, {\it active}), &  \max\{p_1, p_2\} = p_2 > \frac{c}{2} \,.
	\end{cases}
	\end{equation}
	where $\sigma_i$ is any mixed strategy for server $i$. 
\end{theorem} 
~\\
\noindent{\bf Proof:} If $\max\{p_1, p_2\} < \mfrac{c}{2}$ and either server is active, a negative expected payoff is incurred. Thus the best strategy is $( inactive, inactive)$ and the sum of the expected payoffs is $0$.

If $\max\{p_i, p_{-i}\} = p_i > \mfrac{c}{2}$, server $i$ could make its expected payoff positive by choosing to be active, while server $-i$ should remain inactive to avoid an additional cost $c$. In this case, the sum of the expected payoffs is $2p_i-c$. 

If $\max\{p_i, p_{-i}\} = p_i = \mfrac{c}{2}$, server $i$ is indifferent between being inactive and being active, and any mixed strategy $\sigma_i$ results in the same expected payoff of 0. After simple calculations and comparisons, we obtain the results in Equation (\ref{s*}).
\hfill $\Box$

The best strategy $s^*$ is shown in Figure~\ref{1}. The social welfare under strategy $s^*$ is 
\begin{align*}
S_O(s^*) &= \int_{0}^{\frac{c}{2}} \int_{0}^{\frac{c}{2}} 0 \, dp_1 \, dp_2 +  \, \int_{\frac{c}{2}}^{1} \int_{0}^{p_1}  (2p_1 - c) \, dp_2 \, dp_1 +  \, \int_{\frac{c}{2}}^{1} \int_{0}^{p_2}  (2p_2 - c) \, dp_1 \, dp_2  \\
&= \frac{1}{12} \, c^3 -c +\frac{4}{3} \,. \numberthis
\end{align*}
This is the best social welfare under an operating cost $c$ and a uniform distribution for both servers' probabilities of successful transmission. 


\section{Case \uppercase\expandafter{\romannumeral2}: Uncooperative servers without communication}

Case \uppercase\expandafter{\romannumeral2} studies the scenario in which server $i$ is informed that server $-i$ is using a threshold strategy such that it will be active if and only if $p_{-i} \geq p_{-i}^*$. In this section, we first calculate the Nash equilibrium strategy and the social welfare for the values $p_{i}^*$ and $p_{-i}^*$ of the thresholds. We then propose a regulation and conclude that the social welfare under the threshold strategy is maximised by this regulation. 

\subsection{Nash equilibrium strategy in {Case} \uppercase\expandafter{\romannumeral2}}
Guglielmi and Badia\cite{AGLB15} stated that for a given $p_{-i}^*$, if server $i$ is active when its probability of successful transmission is $p_i$, then it should also be active for any other $\tilde{p_i}>p_i$. In this section, we prove that the best response for server $i$ is a threshold strategy. In addition, we calculate a Nash equilibrium strategy and prove this is the only one. This is expressed in Theorem \ref{thm2}.
~\\
\begin{theorem} \label{thm2}
	For $i = 1,2$, server $i$'s best response is of threshold type regardless of server $-i$'s strategy and $p_{-i}$'s distribution. Assume server $i$ knows that server $-i$ is active if and only if $p_{-i} \geq p^*_{-i}$, server $i$ knows $p^*_{-i}$ and that $p_{-i}$ follows a uniform distribution on $[0, 1]$, then server $i$'s best response is also a threshold strategy with 
	\begin{equation}
	BR(p^*_{-i
	}) = 
	\begin{cases}
	\sqrt{2c-p^{*2}_{-i}} \,, & if \, p^*_{-i} \leq \sqrt{c} \\[+6pt]
	\displaystyle{\frac{c}{p^*_{-i}}} \,, &if \, p^*_{-i} > \sqrt{c} \,.  \label{eq1}
	\end{cases}
	\end{equation}
	~\\
	Also, the unique Nash equilibrium strategy is $(\underline{\sqrt{c}}, \underline{\sqrt{c}})$ which means both servers adopt a threshold strategy with $p^*_1 =p^*_2=\sqrt{c}$ .
\end{theorem}

\noindent{\bf Proof:} Assume that server $2$ is adopting any strategy $s_2$. For server $1$, when it is active, its payoff is either $p_1-c$ or $\max \{p_1, p_2\}-c$ when server $2$ is inactive or active, respectively. Both $p_1$ and $\max \{p_1, p_2\}$ are increasing with $p_1$, thus the expected payoff of server $1$ when it is active is increasing with $p_1$. When server $1$ is inactive, its expected payoff is not a function of $p_1$. thus, if server $1$'s best response is to be active at $p_1$, then this is also its best response at any $\tilde{p_1} \geq p_1$. So the best response of server $1$ is of threshold type no matter what strategy the other server adopts. An identical analysis holds for server $2$. In the Nash equilibrium, each server plays a best response against the other server simultaneously, so the Nash equilibrium strategy is of threshold type.

Next, we assume that server $i$ knows its own $p_i$, and that server 2 adopts a strategy with threshold  $p^*_2$, which is known to server $1$. We discuss the expected payoff to server 1 in two cases: $p_1 < p^*_2$ and $p_1 \geq p^*_2$. 

When $p_1 < p^*_2$,
\begin{align*}
&\begin{aligned}
\EX[U_1((active, \underline{p^*_2}),p_1)] &= \int_{0}^{p^*_2}(p_1-c)dp_2 + \int_{p^*_2}^{1} (p_2-c)dp_2  \\
&= \frac{1-p^{*2 }_2}{2} + p_1p^{*}_2 - c \,,
\end{aligned}\\
&\begin{aligned}
\EX[U_1((inactive, \underline{p^*_2}),p_1)] = \int_{p^*_2}^{1}p_2dp_2 = \frac{1-p^{*2}_2}{2} \,.
\end{aligned}
\end{align*}
Then
\begin{equation*}
\EX[U_1((active, \underline{p^*_2}),p_1)]  \geq \EX[U_1((inactive, \underline{p^*_2}),p_1)] \Longleftrightarrow p_1 \in \left[\frac{c}{p^*_2}, p^*_2\right).
\end{equation*}
When $p_1 \geq p^*_2$,
\begin{align*}
&\begin{aligned}
\EX[U_1((active, \underline{p^*_2}),p_1)] &= \int_{0}^{p^*_2}(p_1-c)dp_2 + \int_{p^*_2}^{p_1}(p_1-c)dp_2 + \int_{p_1}^{1} (p_2-c)dp_2 \\
&= \frac{p_1^2+1}{2} - c \,,\\
\end{aligned}\\
&\EX[U_1((inactive, \underline{p^*_2}),p_1)] = \int_{p^*_2}^{1}p_2dp_2 = \frac{1-p^{*2}_2}{2} \,.
\end{align*}
Then 
\begin{equation*}
\EX[U_1((active, \underline{p^*_2}),p_1)]  \geq \EX[U_1((inactive, \underline{p^*_2}),p_1)] \Longleftrightarrow p_1 \in \left[\max\{p^*_2, \sqrt{2c-p^{*2}_2}\}, 1\right].
\end{equation*}

In summary, if $0 \leq p^*_2\leq \sqrt{c}$, then $\left[\displaystyle{\frac{c}{p_2^*}}, p_2^*\right)$ is empty and server 1 never becomes active when $p_1 < p^*_2$. Furthermore,  $\max\{p_2^*,\sqrt{2c-p^{*2}_2}\}= \sqrt{2c-p^{*2}_2}$, so server 1 becomes active when $p_1 \in [\sqrt{2c-p^{*2}_2}, 1]$. On the other hand, if $\sqrt{c} < p^*_2 \leq 1 $, then  $\displaystyle{\frac{c}{p_2^*}} < p_2^*$ and  $\max\{p_2^*,\sqrt{2c-p^{*2}_2}\}= p_2^*$, with the result that $p_1 \in \left[\displaystyle{\frac{c}{p^*_{2}}}, 1\right]$. Thus the best response for server 1 is threshold type with threshold value
\begin{equation}
BR_1(p^*_2) = \begin{cases}
\sqrt{2c-p^{*2}_2}, &  0 \leq p^*_2 \leq \sqrt{c} \\[+6pt]
\displaystyle{\frac{c}{p^*_{2}}}, & \sqrt{c} < p^*_2 \leq 1 \,.
\end{cases}
\end{equation}
which is the same as (\ref{eq1}) when $i=1$. An identical analysis holds when $i=2$.

For the Nash equilibrium strategy, we notice that when $p^*_{-i} \in [0, \sqrt{c}]$, 
\begin{equation*}
p^*_i= \sqrt{2c-p^{*2}_{-i}} \geq \sqrt{c},
\end{equation*}
thus a Nash equilibrium should also satisfy 
\begin{equation*}
p^*_{-i} = \displaystyle{\frac{c}{p^*_i}}.
\end{equation*}
By solving the two equations simultaneously, we obtain the threshold type Nash equilibrium strategy $(\underline{\sqrt{c}},\underline{\sqrt{c}})$; when $p^*_{-i} \in [\sqrt{c}, 1]$, the analysis is similar, and $p^*_i = \mfrac{c}{p^*_{-i}}\in [0, \sqrt{c}]$, thus $p^*_{-i}= \sqrt{2c-p^{*2}_{i}}$. The resulting Nash equilibrium strategy is also  $(\underline{\sqrt{c}},\underline{\sqrt{c}})$.  Thus $(\underline{\sqrt{c}},\underline{\sqrt{c}})$ is the unique Nash equilibrium strategy, that is, each server chooses to be active if and only if its successful transmission probability is greater than or equal to $\sqrt{c}$. \hfill $\Box$

\subsection{Social welfare of uncooperative servers without communication}

\begin{figure}
	\centering
	\includegraphics[width = .45\textwidth]{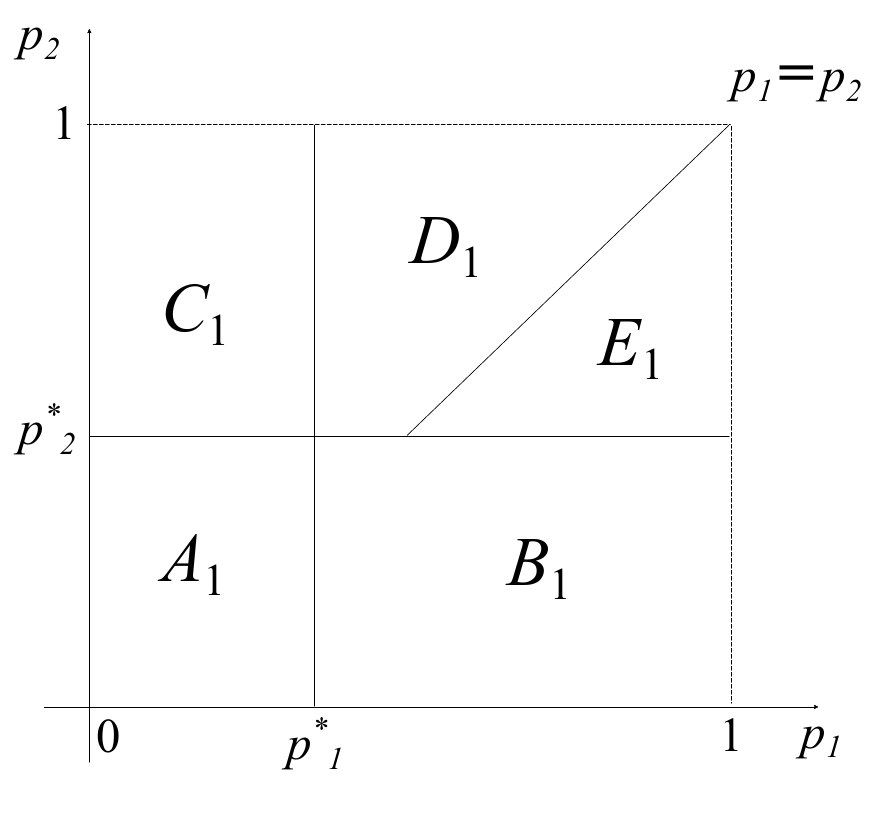}
	\caption{Areas for expected payoff computation when ${p^*_1 < p^*_2}$.} \label{fig2-1}
\end{figure}

\begin{figure}
	\centering
	\includegraphics[width = .45\textwidth]{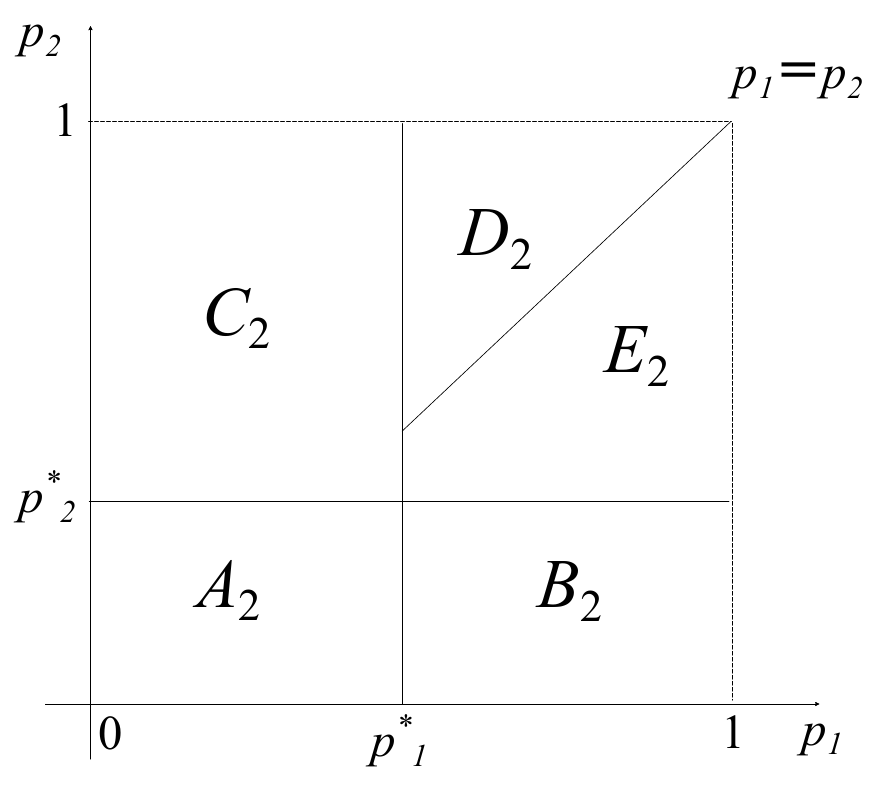}
	\caption{Areas for expected payoff computation when ${p^*_1\geq p^*_2}$.} \label{fig2-2}
\end{figure}

We assume both servers adopt threshold strategies of $p^*_1$ and $p^*_2$ and  discuss the social welfare for the two cases: $p^*_1 < p^*_2$ and $p^*_1 \geq p^*_2$. We denote the social welfare gained from server 1 and server 2 by $T_{1}(\underline{p^*_1},\underline{p^*_2})$ and $T_{2}(\underline{p^*_1},\underline{p^*_2})$ if they adopt threshold strategy $(\underline{p^*_1},\underline{p^*_2})$, respectively. Our computation is based on the partitioned regions shown in Figure~\ref{fig2-1} when $p^*_1 < p^*_2$,  and Figure~\ref{fig2-2} when $p^*_1 \geq p^*_2$. 

When $p^*_1 < p^*_2$, in region $A_1$, both servers are inactive, and both expected payoffs are 0. In region $B_1$, server 1 is active and server 2 is inactive, and the expected payoffs of server 1 and server 2 are $p_1-c$ and $p_1$, respectively. Region $C_1$ is similar to $B_1$, server 1 is inactive and server 2 is active, and the expected payoffs to server 1 and server 2 are $p_2$ and $p_2-c$, respectively. In $D_1$ and $E_1$, both servers are active. In $D_1$ where $p_1 \leq p_2$, both expected payoffs are $p_2 - c$, while in $E_1$ where $p_1 > p_2$, both expected payoffs are $p_1 - c$. The expected payoffs for server 1 in the five regions in Figure~\ref{fig2-1} and Figure~\ref{fig2-2} are
\begin{align*}
A_1 &= \int_{0}^{p^*_1}\int_{0}^{p^*_2} 0 \, dp_2 \, dp_1  = 0 \\
B_1 &= \int_{p^*_1}^{1}\int_{0}^{p^*_2} (p_1-c)\,dp_2 \, dp_1 = \left( \frac{1}{2}-c \, (1-p^*_1)-\frac{p^{*2}_1}{2} \right) \, p^*_2\\
C_1 &= \int_{0}^{p^*_1}\int_{p^*_2}^{1} p_2 \, dp_2\,dp_1 = p^*_1 \, \left( \frac{1}{2} - \frac{p^{*2}_2}{2} \right) \\
D_1 &= \int_{p^*_2}^{1}\int_{p^*_1}^{p_2} (p_2-c) \, dp_1 \, dp_2 = \frac{1}{3} - \frac{c}{2} - p^*_1 \, \left( \frac{1}{2} - c + c \, p^*_2 \right) + \left( \frac{c}{2} + \frac{p^*_1}{2} - \frac{p^*_2}{3} \right) \, p^{*2}_2 \\
E_1 &= \int_{p^*_2}^{1}\int_{p_2}^{1} (p_1-c) dp_1dp_2 = \frac{1}{6} \, \left( p^*_2-3c+2 \right) \, \left( p^*_2-1 \right)^2 \,. 
\end{align*}
Thus the expected payoff for server 1 if $p^*_1 < p^*_2$ is
\begin{equation*}
T_1(\underline{p^*_1},\underline{p^*_2}) = A_1 + B_1 +C_1 +D_1+E_1 = \frac{1}{6} \, \left( 4+6 \, c \, \left(p^*_1-1 \right) -3 \, p^{*2}_1 \, p^*_2 - p^{*3}_2\right) \,.
\end{equation*}
A similar procedure follows for server 2. Server 2's expected payoff if $p^*_1 < p^*_2$ is
\begin{equation*}
T_2(\underline{p^*_1},\underline{p^*_2}) =\frac{1}{6} \, \left( 4+6 \, c \, \left(p^*_2-1 \right) -3 \, p^{*2}_1 \, p^*_2 - p^{*3}_2\right) \,.
\end{equation*}
Thus, the social welfare 
\begin{equation*}
S_O(\underline{p^*_1},\underline{p^*_2}) = T_1(\underline{p^*_1},\underline{p^*_2}) + T_2(\underline{p^*_1},\underline{p^*_2}) = \frac{4}{3} + c \, \left( p^*_1 + p^*_2 - 2  \right) - \frac{1}{3} \, \left( 3 \, p^{*2}_1 + p^{*2}_2 \right) \, p^{*}_2 \,.
\end{equation*}
%
%
When $p^*_1 \geq p^*_2$, by symmetry, 
%
the social welfare 
\begin{equation*}
S_O(\underline{p^*_1},\underline{p^*_2}) = T_1(\underline{p^*_1},\underline{p^*_2}) + T_2(\underline{p^*_1},\underline{p^*_2}) =  \frac{4}{3} + c \, \left( p^*_1 + p^*_2 - 2  \right) - \frac{1}{3} \, p^{*}_1 \, \left( p^{*2}_1 + 3 \, p^{*2}_2 \right)  \,.
\end{equation*}
${S}_O(\underline{p^*_1},\underline{p^*_2})$ attains its maximum at $\left(\underline{\sqrt{\mfrac{c}{2}}}, \underline{\sqrt{\mfrac{c}{2}}}\right)$, while the unique Nash equilibrium strategy is $(\underline{\sqrt{c}}, \underline{\sqrt{c}})$ and 
\begin{align*}
&S_O \left(\underline{\sqrt{\mfrac{c}{2}}}, \underline{\sqrt{\mfrac{c}{2}}}\right) = \frac{2\sqrt{2}}{3} \, c \sqrt{c} -2 \, c +\frac{4}{3}\\
&S_O (\underline{\sqrt{c}}, \underline{\sqrt{c}}) = \frac{2}{3} \, c \sqrt{c} -2 \, c +\frac{4}{3}\,.
\end{align*}
Therefore, $S_O (\underline{\sqrt{c}}, \underline{\sqrt{c}})  \leq S_O \left(\underline{\sqrt{\mfrac{c}{2}}}, \underline{\sqrt{\mfrac{c}{2}}}\right)$. Hence $(\underline{\sqrt{c}}, \underline{\sqrt{c}})$ is not the social optimal threshold strategy.


If we impose a regulation that whichever server chooses to be active, the other server pays it $\mfrac{c}{2}$ as a subsidy, then the Nash equilibrium becomes the socially optimal solution. For example, if server 1 is active while server 2 is inactive, the previous payoff $(p_1 - c, p_1)$ becomes $\left(p_1 - \mfrac{c}{2}, p_1 - \mfrac{c}{2}\right)$; if both servers are active, the payoffs remain unchanged. The regulation is essentially making the original game fairer. If we use $\tilde{\mathbb{E}}[U_i((active , \underline{p^*_{-i}}), p_i)]$ to denote the expected payoff of server $i$ after regulation, then when $p_1 < p^*_2$,
\begin{align*}
&\begin{aligned}
\tilde{\mathbb{E}}[U_1((active, \underline{p^*_2}),p_1)] &= \int_{0}^{p^*_2}(p_1-c)dp_2 + \int_{p^*_2}^{1} (p_2-c)dp_2 + \int_{0}^{p^*_2}\left(\frac{c}{2}-c\right)dp_2 \\
&= \frac{1-p^{*2 }_2}{2} + p_1p^{*}_2  + \frac{c}{2}\,p^*_2  - c
\end{aligned}\\
&\begin{aligned}
\tilde{\mathbb{E}}[U_1((inactive, \underline{p^*_2}),p_1)] = \int_{p^*_2}^{1}\left(p_2-\frac{c}{2}\right)dp_2 = \frac{1-p^{*2}_2}{2} - \frac{c}{2}\,(1-p^*_2 )\,,
\end{aligned}
\end{align*}
hence
\begin{equation*}
\tilde{\mathbb{E}}[U_1((active, \underline{p^*_2}),p_1)] \geq \tilde{\mathbb{E}}[U_1((inactive, \underline{p^*_2}),p_1)] \Longleftrightarrow p_1 \in \left[\frac{c}{2\,p^*_2}, p^*_2\right).
\end{equation*}
When $p_1 \geq p^*_2$,
\begin{align*}
&\begin{aligned}
\tilde{\mathbb{E}}[U_1((active, \underline{p^*_2}),p_1)] &= \int_{0}^{p^*_2}p_1dp_2 + \int_{p^*_2}^{p_1}p_1dp_2 + \int_{p_1}^{1} p_2dp_2 + \int_{0}^{p^*_2}\frac{c}{2}dp_2 - c \\
&= \frac{1+p_1^2}{2} + \frac{c}{2}\,p^*_2 -c 
\end{aligned}\\
&\begin{aligned}
\tilde{\mathbb{E}}[U_1((inactive, \underline{p^*_2}),p_1)] = \int_{p^*_2}^{1}\left(p_2-\frac{c}{2}\right)dp_2 = \frac{1-p^{*2}_2}{2} - \frac{c}{2}\,(1-p^*_2 )\,,
\end{aligned}
\end{align*}
hence
\begin{equation*}
\tilde{\mathbb{E}}[U_1((active, \underline{p^*_2}),p_1)] \geq \tilde{\mathbb{E}}[U_1((inactive, \underline{p^*_2}),p_1)] \Longleftrightarrow p_1 \in \left[\max\{p^*_2, \sqrt{c-p^{*2}_2}\}, 1\right].
\end{equation*}
Thus
\begin{align*}
& \tilde{\mathbb{E}}[U_1((active, \underline{p^*_2}),p_1)] \geq \tilde{\mathbb{E}}[U_1((inactive, \underline{p^*_2}),p_1)] \\\\[+2pt]
& \Longleftrightarrow p_1 \in \left[\frac{c}{2\,p^*_2}, p^*_2\right] \, \cup \, \left[\max\{p^*_2, \sqrt{c-p^{*2}_2}\}, 1\right] \\\\[+2pt]
& \Longleftrightarrow BR_1(p^*_2)\in
\begin{cases}
\sqrt{c-p^{*2}_2}, &  0 \leq p^*_2 \leq \sqrt{\mfrac{c}{2}} \\[+6pt]
\displaystyle{\frac{c}{2\,p^*_{2}}}, & \sqrt{\mfrac{c}{2}} < p^*_2 \leq 1 \,.
\end{cases}
\end{align*}
A similar analysis as in the proof of Theorem \ref{thm2} establishes that $\left(\underline{\sqrt{\mfrac{c}{2}}}, \underline{\sqrt{\mfrac{c}{2}}},\right)$ is the unique Nash equilibrium strategy. 

We remark here that if $p_1$ and $p_2$ independently follow the same distribution on $[0, 1]$ whose cumulative distribution function is $F(\cdot)$, then the threshold Nash equilibrium strategy is unique and is of type $(\underline{h(c)}; \underline{h(c)})$, where $h(c)$ satisfies $h(c)F(h(c)) = c$.

\section{Case \uppercase\expandafter{\romannumeral3}: Uncooperative servers with communication}

Case \uppercase\expandafter{\romannumeral3} analyses the situation where both servers have full information about each other's $p_i$, but each server maximises its own payoff. We first derive the Nash equilibria for the whole state space, and then calculate the best and worst social welfare within the Nash equilibrium strategies. Finally, we apply a regulation and obtain the social welfare based on a regulated Nash equilibrium strategy which is exactly the same as $S_O(s^*)$ in Case \uppercase\expandafter{\romannumeral1}.

\subsection{Nash equilibrium strategy in Case \uppercase\expandafter{\romannumeral3} }

The Nash equilibrium in this case is similar to that of Case \uppercase\expandafter{\romannumeral1}, but both servers are more conservative in their decisions to be active. The Nash equilibrium strategy is expressed in Theorem \ref{thm3}.
\\[1pt]
%

\begin{theorem}\label{thm3}
	If $p = (p_1, p_2)$ is observable to both servers when they make decisions and both servers care only about their own payoffs, then the Nash equilibrium strategy on the whole state space is given by
	\begin{align}
	s^*_N =
	\begin{cases}
	(inactive, inactive), \qquad \quad \, \, \text{if }\max\{p_1, p_2\} < c \\
	(\sigma_1, inactive), \qquad \qquad \quad \, \, \text{if }\max\{p_1, p_2\} = p_1 = c\\
	(inactive, \sigma_2), \qquad \qquad \quad \, \, \text{if }\max\{p_1, p_2\} = p_2 = c\\
	(inactive, active)\,, \qquad \qquad \text{if } \{p_1 < c < p_2\} \cup \{p_2-p_1 > c\} \\
	(active, inactive)\,, \qquad \qquad \text{if }   \{p_2 < c < p_1\} \cup \{p_1-p_2 > c\}  \\
	(inactive, active) \, \textup{or} \, (active, inactive) \, \textup{or} \, \left(\mfrac{p_2-c}{p_2},\mfrac{p_1-c}{p_2}\right)\,, \\
	\quad \quad \quad \text{if }   \{0 \leq p_1-p_2 \leq c\} \cap \{p_1 > c\} \cap \{p_2 > c\} \\
	(inactive, active) \,\textup{or} \, (active, inactive) \, \textup{or} \, \left(\mfrac{p_2-c}{p_1},\mfrac{p_1-c}{p_1}\right)\,,\\
	\quad \quad \quad \text{if }   \{0 < p_2-p_1 \leq c\} \cap \{p_1 > c\} \cap \{p_2 > c\} \,.
	\end{cases}
	\end{align}
	where $\sigma_i$ is any mixed strategy for server $i$. 
\end{theorem}
~\\
\noindent{\bf Proof:} When $\max\{p_1, p_2\} < c$, {\it inactive} is the dominant strategy for both servers, so $(inactive, inactive)$ is the Nash equilibrium. On the boundary $\max\{p_i, p_{-i}\} = p_i = c$, as server $i$ is indifferent between being active and being inactive, any mixed strategy $\sigma_i$ is a Nash equilibrium and achieves zero payoff.

When $p_2-c >p_1$, then server 2 has a higher expected payoff if it chooses to be active irrespective of what server 1 chooses to do. So server 2 should be active. Then server 1 should choose to be inactive. 

When $p_1<c<p_2$, then, if server 2 is inactive, server 1 has the choice between an expected payoff of 0 if it is inactive and $p_1-c<0$ if it is active. So it should choose to be inactive. Alternatively, if server 2 chooses to be active then server 1 has the choice between expected payoffs of $p_2$ if it is inactive, and $p_2-c$ if it is active. So again it should choose to be inactive. Given that server 1 is inactive, server 2 has a choice between expected payoffs of $p_2-c>0$ if it chooses to be active and 0 if it chooses to be inactive. So it should choose to be active. This gives us $(inactive,active)$ as the optimal strategy if $p_1<c<p_2$. 

Similarly, if $p_2<c<p_1$ or $p_1-p_2>c$, then the optimal strategy is $(active,inactive)$.

When $|p_1-p_2| \leq c$ and $\min\{p_1, p_2\} > c$, if server 1 is active, the expected payoff for server 2 if it is active is $\max\{p_1, p_2\} -c $ which is less than or equal to $p_1$, so the best response for server 2 is to choose to be inactive. On the other hand, if server 1 is inactive, the best response of server 2 is to choose to be active because its expected payoff $p_2-c$ is positive. So the conclusion is that if $|p_1-p_2|\leq c$ and $\min(p_1,p_2)>c$, it is a Nash equilibrium for server 1 to be active precisely when server 2 is inactive and vice versa. This means that any partition of this region into disjoint sets where the strategy is $(active,inactive)$ and $(inactive, active)$ corresponds to a Nash equilibrium strategy. If $p_1 \geq p_2$, when server 1 is active with probability $\mfrac{p_2-c}{p_2}$, the expected payoffs of server 2 when it is active and inactive are both $\mfrac{p_1(p_2-c)}{p_2}$, that is,  server 2 is indifferent among any mixed strategy; when server 2 is active with probability $\mfrac{p_1-c}{p_2}$, the expected payoffs of server 1 when it is active and inactive are both $p_1-c$, so server 1 is indifferent among any mixed strategy. Thus, if the strategy profile $\left(\mfrac{p_2-c}{p_2},\mfrac{p_1-c}{p_2}\right)$ is adopted by both servers, neither side has an incentive to deviate, and we conclude that $\left(\mfrac{p_2-c}{p_2},\mfrac{p_1-c}{p_2}\right)$ is a mixed Nash equilibrium strategy when $p_1 \geq  p_2$. In this case, if server 1's probability to be active changes from $\mfrac{p_2-c}{p_2}$ to a larger value (smaller value), then for server 2, it pays to stay inactive (active); if server 2's probability to be active changes from $\mfrac{p_1-c}{p_2}$ to a larger value (smaller value), then for server 1, it pays to stay inactive (active). Thus, $\left(\mfrac{p_2-c}{p_2},\mfrac{p_1-c}{p_2}\right)$ is not a stable Nash equilibrium. Similarly, $\left(\mfrac{p_2-c}{p_1},\mfrac{p_1-c}{p_1}\right)$ is a mixed Nash equilibrium strategy when $p_1 <  p_2$ and it is not a stable Nash equilibrium either.
\hfill $\Box$

\begin{figure}
	\centering
	\includegraphics[width = .65\textwidth]{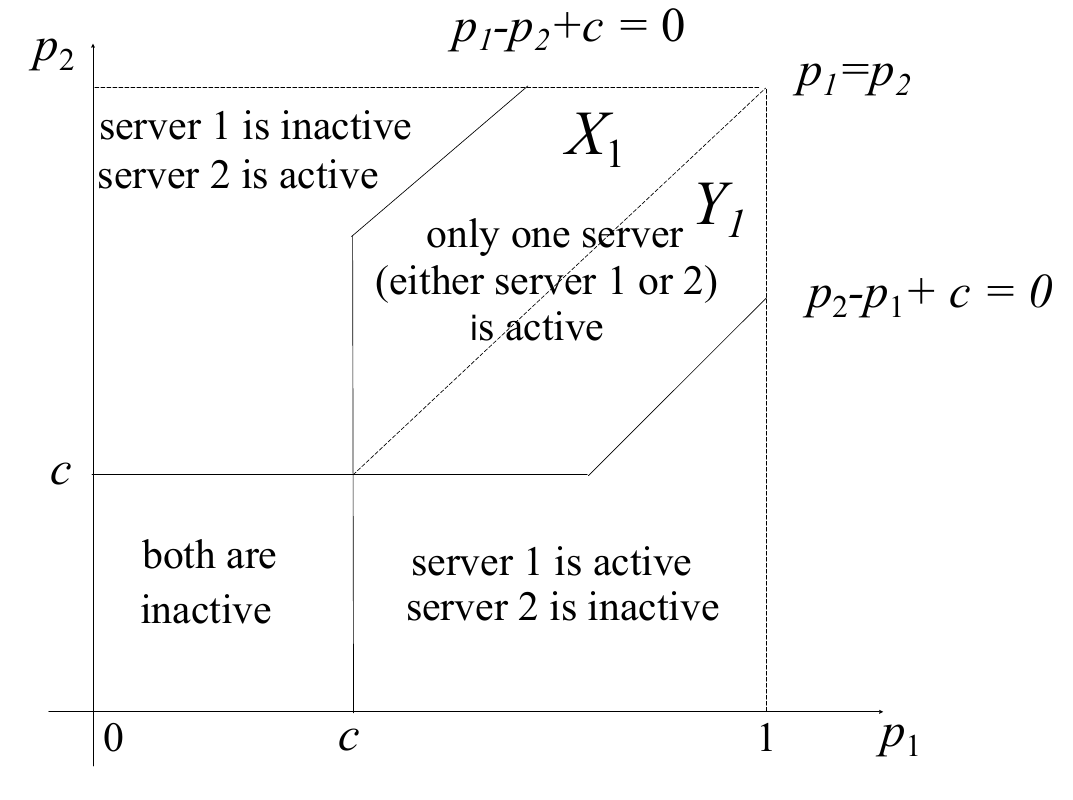}
	\caption{Nash equilibrium for uncooperative servers with communication where ${0 <c < 0.5}$,}\label{fig3-1}
\end{figure}

\begin{figure}
	\centering
	\includegraphics[width = .65\textwidth]{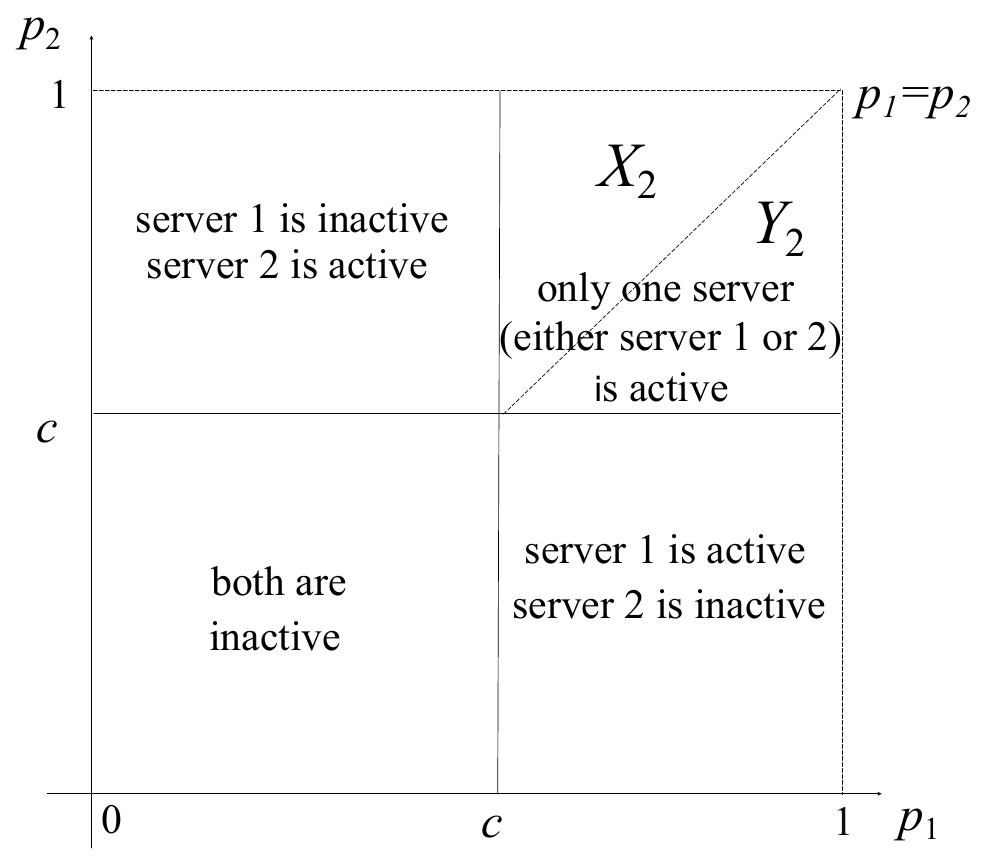}
	\caption{Nash equilibrium for uncooperative servers with communication where ${0.5 \leq c < 1}$.} \label{fig3-2}
\end{figure}

The Nash equilibrium strategy $s^*_N$ is shown in Figure \ref{fig3-1} when $0 <c < 0.5$, and Figure \ref{fig3-2} when $0.5 \leq c < 1$\,. 

\subsection{Social welfare of uncooperative servers with communication}
It follows from the analysis above that there are multiple Nash equilibria corresponding to the set $\{|p_1-p_2| \leq c\} \cap \{p_1 > c\} \cap \{p_2 > c\}$. Any partition of this region into disjoint sets where server 1 is active and server 2 is active results in a Nash equilibrium. The maximum social welfare is obtained if the strategy is $(active, inactive)$ when $\{p_1-p_2 \leq c\} \cap \{p_1 \geq p_2 >  c\} $ and $(inactive, active)$ when $\{p_2-p_1 \leq c\} \cap \{p_2 > p_1 >  c\} $. The minimum social welfare is attained when the strategy is just the opposite which is $(inactive, active)$ when $\{p_1-p_2 \leq c\} \cap \{p_1 \geq p_2 >  c\} $ and $(active, inactive)$ when $\{p_2-p_1 \leq c\} \cap \{p_2 > p_1 >  c\} $.  

When $0 \leq c < 0.5$, the case is shown in Figure \ref{fig3-1}. The maximum social welfare Nash equilibrium strategy is $(inactive,  active)$ in region $X_1$ and $(active, inactive)$ in region $Y_1$. The sum of the expected payoffs is $2p_2-c$  in $X_1$ and $2\,p_1-c$ in $Y_1$. The minimum social welfare Nash equilibrium strategy is $(active, inactive)$ in region $X_1$ and $(inactive, active)$ in region $Y_1$. The sum of the expected payoffs is  $2p_1-c$  in $X_1$ and $2\,p_2-c$ in $Y_1$. By noticing that the sum of the expected payoffs is symmetric with respect to $p_1 = p_2$, the maximum and minimum social welfare within the set of Nash equilibria are
\begin{align*}
\max &S_O(s^*_N) \\
&= \int_{0}^{c} \int_{0}^{c} 0 \, dp_1 \, dp_2 +  \, \int_{c}^{1} \int_{0}^{p_1}  (2p_1 - c) \, dp_2 \, dp_1 + \, \int_{c}^{1} \int_{0}^{p_2}  (2p_2- c) \, dp_1 \, dp_2 \\
&= -\frac{1}{3} \, c^3 - c + \frac{4}{3} \,, \numberthis \\
\textup{and}\\
\min &S_O(s^*_N) = \int_{0}^{c} \int_{0}^{c} 0 \, dp_1 \, dp_2 + 2 \, \left(\int_{0}^{c}\int_{c}^{2c} (2 \, p_2 -c) dp_2 \, dp_1 \right.\\
& + \int_{2c}^{1}\int_{0}^{p_2-c} (2 \, p_2 - c) \, dp_1 \, dp_2 + \int_{c}^{1-c}\int_{p_1}^{p_1+c} (2 \, p_1 - c) \, dp_2 \, dp_1 \\
&\quad \left.+ \int_{1-c}^{1}\int_{p_1}^{1} (2 \, p_1 - c) \, dp_2 \, dp_1 \right) = 3\, c^3-2\, c^2 -c +\frac{4}{3} \,, 
\numberthis
\end{align*} 
respectively.

When $0.5 \leq c < 1$, the case is shown in Figure \ref{fig3-2}. The maximum social welfare Nash equilibrium strategy is $(inactive, active)$ in region $X_2$ and $(active, inactive)$ in region $Y_2$. The sum of the expected payoffs is $2p_2-c$  in $X_2$ and $2\,p_1-c$ in $Y_2$. The minimum social welfare Nash equilibrium strategy is $(active, inactive)$ in region $X_2$ and $(inactive, active)$ in region $Y_2$. The sum of the expected payoffs is $2p_1-c$  in $X_2$ and $2\,p_2-c$ in $Y_2$. As above, by noticing that the sum of the expected payoffs is symmetric with respect to $p_1 = p_2$, the maximum and minimum social welfare within the set of Nash equilibria are
\begin{align*}
& \max S_O(s^*_N) \\
& = \int_{0}^{c} \int_{0}^{c} 0 \, dp_1 \, dp_2  +  \, \int_{c}^{1} \int_{0}^{p_1}  (2p_1 - c) \, dp_2 \, dp_1 + \, \int_{c}^{1} \int_{0}^{p_2}  (2p_2- c) \, dp_1 \, dp_2 \\
& = -\frac{1}{3} \, c^3 - c + \frac{4}{3} \, {,}\numberthis \\ 
\textup{and}\\
&\min S_O(s^*_N) \\
& = \int_{0}^{c} \int_{0}^{c} 0 \, dp_1 \, dp_2 + 2 \, \left(\int_{0}^{c}\int_{c}^{1} (2 \, p_2 -c) \, dp_2 \, dp_1 \right.\left.+ \int_{c}^{1} \int_{c}^{p_2} (2 \, p_1 -c) \, dp_1 \, dp_2\right)\\
&= \frac{1}{3} c^3-2 \, c^2+c+\frac{2}{3} \textup{{,}} \numberthis 
\end{align*}
respectively.

We compare the action profile of the maximum social welfare with the social optimum strategy in Case \uppercase\expandafter{\romannumeral1}. When $p_1\leq c$ and $p_2\leq c$, uncooperative servers choose to be inactive, but once $\max\{p_1, p_2\} > \mfrac{c}{2}$, the sum of both servers' expected payoffs is positive, thus, from the system manager's point of view, the server with the higher successful transmission probability should be active. The reason why it stays inactive until its probability of successful transmission exceeds $c$ is because of the unfair division of the total payoff. If the server chooses to be active, it will gain a negative expected payoff while the other server will get an advantage.

We suggest imposing a regulation where the inactive server gives the active server $c-\mfrac{p_1+p_2}{2}$ if $\min\{p_1, p_2\} > \mfrac{c}{2}$. Then when $\max\{p_1, p_2\} < \mfrac{c}{2}$, the game remains unchanged, so the Nash equilibrium is still $(inactive, inactive)$; when $\max\{p_1, p_2\} \geq \mfrac{c}{2}$, the game has different expected payoffs as shown in Table \ref{tab2}.
\begin{table}
	\setlength{\extrarowheight}{2pt}
	\centering
	\begin{tabular}{cc|c|c|}
		& \multicolumn{1}{c}{} & \multicolumn{2}{c}{server $2$}\\
		& \multicolumn{1}{c}{} & \multicolumn{1}{c}{$active$}  & \multicolumn{1}{c}{$inactive$} \\\cline{3-4}
		\multirow{3}*{server $1$}  & $ $ & $\max\{p_1,p_2\} - c, $ & $ \frac{p_1-p_2}{2}, $ \\
		& $active$ & $\max\{p_1,p_2\} - c$ & $ \frac{3\,p_1+p_2}{2}-c$ \\\cline{3-4}
		& $ $ & $ \frac{p_1+3\,p_2}{2}-c,$ & $ $ \\
		& $inactive$ & $  \frac{p_2-p_1}{2}$ & $0,0$ \\\cline{3-4} 
	\end{tabular}	
	\caption{Expected payoff matrix with regulation when ${\max\{p_1, p_2\} \geq \ds\frac{c}{2}}$}\centering\label{tab2}
\end{table}
If $p_1 > p_2$, since $\mfrac{p_1-p_2}{2} > 0$ and $p_1-c+\mfrac{p_1+p_2}{2} > p_1-c$, but $\mfrac{p_2-p_1}{2} < 0$, so $(active, inactive)$ is the only Nash equilibrium. If $p_1 < p_2$, $\mfrac{p_2-p_1}{2} > 0$ and $p_2-c+\mfrac{p_1+p_2}{2} > p_2-c$, but $\mfrac{p_1-p_2}{2} < 0$, so $(inactive, active)$ is the only Nash equilibrium. If $p_1 = p_2$, both $(active, inactive)$ and $(inactive, active)$ are Nash equilibria. After this regulation is imposed, the resulting Nash equilibrium strategy is exactly the same as $s^*$ in Case \uppercase\expandafter{\romannumeral1}, which means that the regulation eliminates the effect of noncooperation. 


\section{Conclusion}

In this paper we have quantified the value of communication and cooperation, and proposed regulation to increase social welfare by eliminating the loss due to noncooperation in a service system with two strategic servers proposed by Guglielmi and Badia \cite{AGLB15}. 

We have applied game theory to analyse the behaviour of the servers where both servers (I) know each other's $p_i$ and they cooperate to maximise social welfare; (II) do not know each other's $p_i$, but each server knows that the other adopts a threshold strategy and what the threshold value is; (III) know each other's $p_i$ but they only maximise their own expected payoff. We computed Nash equilibrium strategy for Cases II and III, and showed that the unique Nash equilibrium strategy in Case II is $(\underline{\sqrt{c}}, \underline{\sqrt{c}})$, that is, each server will be inactive until its probability of successful transmission is at least $\sqrt{c}$. Furthermore we observed that there are multiple Nash equilibria for Case III.

\begin{figure}
	\centering
	\includegraphics[width = .9\textwidth]{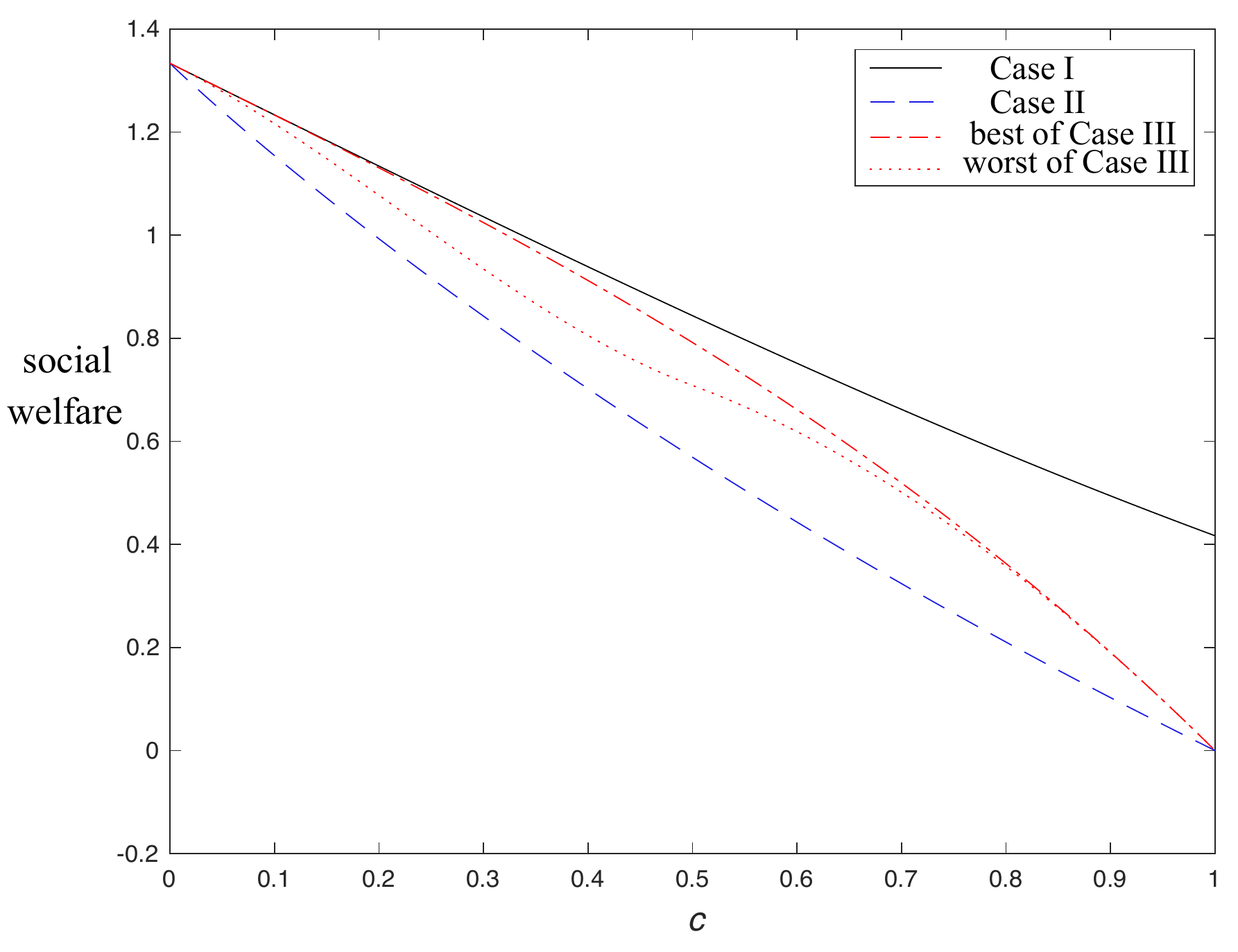}
	\caption{Social welfare under Cases I, II, and III without regulation} \centering\label{c1}
\end{figure}

\begin{figure}
	\centering
	\includegraphics[width = .9\textwidth]{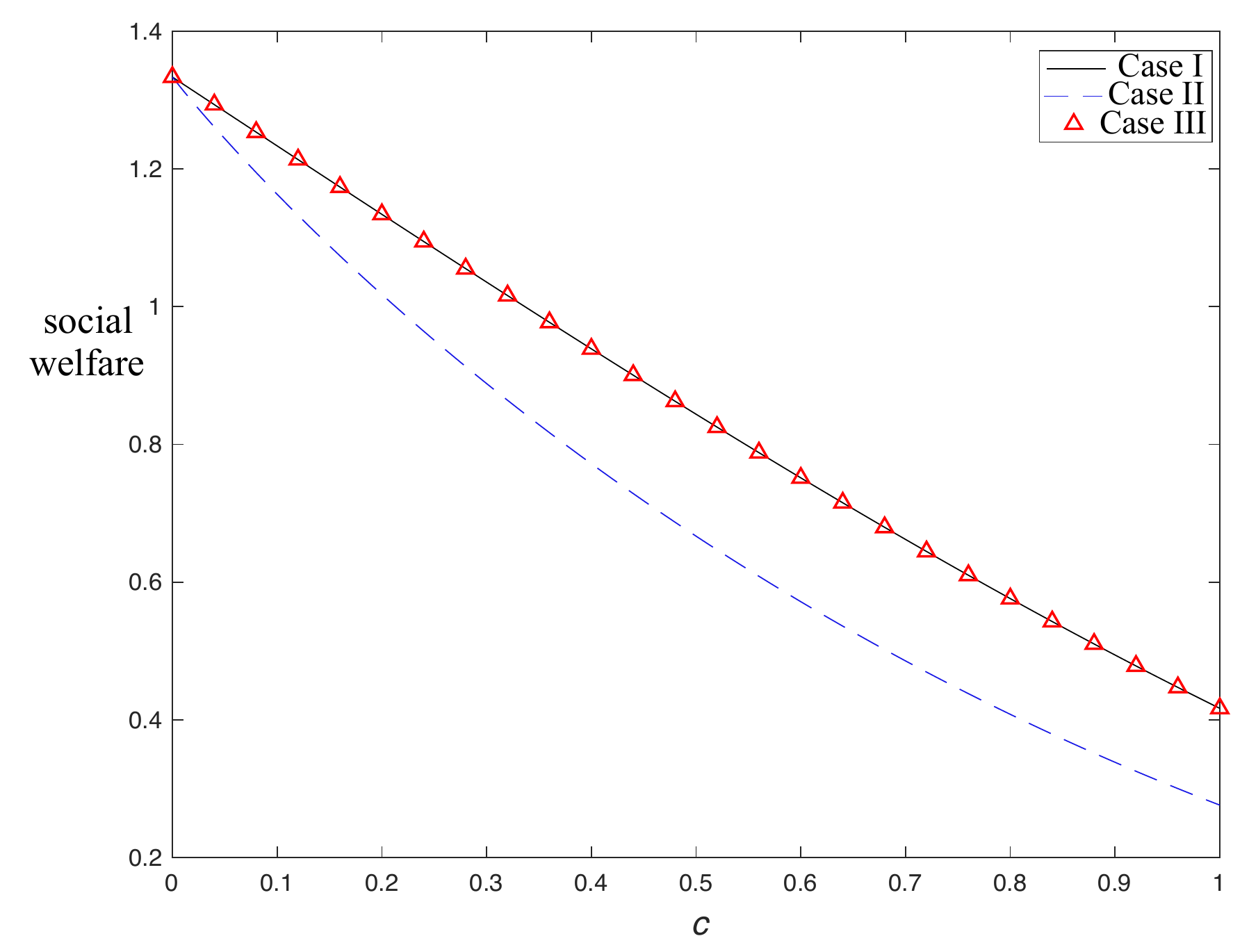}
	\caption{Social welfare under Cases I, II, and III with  regulation} \centering\label{c2}
\end{figure}

We compared the social welfare of the Nash equilibrium strategies in the three cases plotted in Figure~\ref{c1}. We showed that the social welfare for Case II is the worst. This is reasonable since in Case II, servers lack both communication and cooperation. The best result of Case III is still less than that of Case I. This is caused by noncooperation. After imposing regulation, the Nash equilibrium strategy of Case III can be increased so that its expected payoff is the same as that obtained under the socially optimal strategy. For Case II, due to the lack of communication, there is still a gap between the social welfare and the social optimum, but the maximum is attained under the assumption that both servers adopt threshold strategies. The social welfare of the three cases after regulation is plotted in Figure~\ref{c2}. 

\paragraph{Acknowledgements}
\noindent P. G. Taylor's research is supported by the Australian Research Council (ARC) Laureate Fellowship FL130100039 and the ARC Centre of Excellence for the Mathematical and Statistical Frontiers (ACEMS). M. Fackrell's research is supported by the ARC Centre of Excellence for the Mathematical and Statistical Frontiers (ACEMS). J. Wang would like to thank the University of Melbourne for supporting her work through the Melbourne Research Scholarship. The authors are also grateful to two anonymous referees, whose enlightened comments helped improve our work.



%
%
%


\end{document}